\documentclass[aps,prd,twocolumn,showpacs,preprintnumbers,nofootinbib]{revtex4}
\usepackage{amssymb,latexsym}
\usepackage{amsmath,amsbsy}
\usepackage{epsfig,bm}
\usepackage{feynmf}
\DeclareMathOperator{\pplus}{p^{+}}

\DeclareMathOperator{\kplus}{k^{+}}

\DeclareMathOperator{\ie}{i \epsilon}
\DeclareMathOperator{\Res}{Res}

\DeclareMathOperator{\dw}{D_{\text{W}}}

\DeclareMathOperator{\kminus}{k^{--}}
\DeclareMathOperator{\pminus}{p^{--}}

\DeclareMathOperator{\Gt}{\tilde{G}}

\DeclareMathOperator{\xpp}{x^{\prime\prime}}
\DeclareMathOperator{\yp}{y^\prime}

\begin{document}
\preprint{NT-UW 02-30}
\title{Current in the light-front Bethe-Salpeter formalism I:\\ Replacement of non-wave function vertices}
\author{B.~C.~Tiburzi}
%\email{bctiburz@u.washington.edu}
\author{G.~A.~Miller}
%\email{miller@phys.washington.edu}
\affiliation{Department of Physics  
	University of Washington      
	Box 351560
	Seattle, WA 98195-1560}
\date{\today}

\begin{abstract}
We apply the light-front reduction of the Bethe-Salpeter equation to matrix elements of the electromagnetic current
between bound states. Using a simple $(1+1)$-dimensional model to calculate form factors, we focus on two cases. 
In one case, the interaction is dominated by a term instantaneous in light-cone time. 
Here effects of higher Fock states are negligible and the form factor
can be effectively expressed using non-wave function vertices and crossed interactions. 
If the interaction is not instantaneous, non-wave function
vertices are replaced by contributions from higher Fock states. These higher  
Fock components arise from the covariant formalism via the energy poles of the Bethe-Salpeter vertex
and the electromagnetic vertex. 
%Contributions to covariant constituent quark models from instantaneous and
%non-wave function vertices are also investigated. 
The replacement of non-wave function vertices in time-ordered
perturbation theory is a theorem which directly extends to generalized parton distributions, e.g., 
in $(3+1)$ dimensions. 
\end{abstract}

\pacs{11.10.St, 11.40.-q, 13.40.Gp}

\maketitle

\section{Introduction}
More than a half century ago, Dirac's paper on the forms of relativistic dynamics \cite{Dirac:1949cp} introduced the front-form
Hamiltonian approach.  Applications to quantum mechanics and field theory were overlooked at the time due to the 
appearance of covariant perturbation theory.
%, in which the special role played by time was supplanted by Lorentz invariance. 
%Front-form dynamics gradually resurfaced, first under the guise of the infinite momentum frame (in which covariant calculations
%could be achieved more easily if the processes were viewed at the speed of light) and later as quantization on the light-front plane
%(or null plane). 
The reemergence of front-form dynamics was largely motivated by simplicity as well as physicality. The 
light-front approach has the largest stability group \cite{Leutwyler:1977vy} of any Hamiltonian theory.
%, that is light-front dynamics
%maintains the smallest number of interaction dependent generators of the Poincar\'e algebra. In highly relativistic calculations the utility is clear:
%Lorentz boosts on the light front are merely kinematical; whereas in the instant form, boosts depend on the dynamical evolution of the system. 
%The physical motivation for hard scattering processes, such as deeply inelastic scattering, was the infinite momentum frame's utility 
%in describing partonic subprocesses. 
Today the physical connection to light-front dynamics is transparent: hard scattering processes probe 
a light-cone correlation of the fields. Not surprisingly, then, many perturbative QCD applications can be treated on the light front, see e.g. 
\cite{Lepage:1980fj}. Outside this realm, physics on the light cone has been extensively developed for non-perturbative QCD 
\cite{Brodsky:1997de} as well as applied to nuclear physics \cite{Carbonell:1998rj,Miller:2000kv}.

This paper concerns current matrix elements between bound states of two particles in the light-front formalism. We approach the topic,
however, from covariant perturbation theory. 
%By projecting covariant quantities onto the light cone, one has a way of deriving
%light-front amplitudes from field theory, circumventing the often subtle task of formulating the front form of dynamics \emph{ab initio}.
As demonstrated by the tremendous undertaking of \cite{Ligterink:1994tm}, one can derive light-front perturbation theory for scattering states
by projecting covariant perturbation 
theory onto the light cone, thereby demonstrating their equivalence---including the delicate issue of renormalization. As to 
the issue of light-front bound states, a reduction scheme for the Bethe-Salpeter equation recently appeared \cite{Sales:1999ec} 
that produces a kernel calculated in light-front perturbation theory. For the purpose of simplicity, we consider only bound states of 
two scalars interacting via the exchange of a massive scalar in the $(1+1)$-dimensional ladder model. This work supersedes our original 
investigation \cite{Tiburzi:2002mn}.

Our main consideration is to extend the reduction to current matrix elements to investigate valence and non-valence contributions in the 
light-front reduced, Bethe-Salpeter formalism. Thus we calculate our model's form factor; moreover,  in $(1+1)$ dimensions we cannot 
choose a frame of reference where Z-graphs vanish. This enables us to completely investigate their contribution, 
which in $(3+1)$ dimensions has a variety of applications such as generalized parton distributions \cite{Ji:1998pc}. These applications
are pursued elsewhere \cite{future}. Z-graph contributions haunt 
light-front dynamics since non-valence properties of the bound state are involved, so that 
valence wave function models cannot be utilized directly.  On one hand, the light-cone Fock representation provides expressions for the 
Z-graph contributions in terms of Fock component overlaps which are non-diagonal in particle number \cite{Brodsky:2001xy}. 
While on the other hand, vertices which cannot be related to the valence wave function (coined as \emph{non-wave function vertices} 
in \cite{Bakker:2000rd}) 
appear in the light-front Bethe-Salpeter formalism. A variety of ways have been proposed for dealing with these non-wave function vertices
\cite{Einhorn:1976uz,Ji:2000fy,Tiburzi:2001je,Hwang:2001wu}. We also note that one can avoid the issue by attempting to model the 
covariant vertex \cite{deMelo:1997cb,Jaus:zv}, or (when possible) estimating the contribution from higher Fock states \cite{Demchuk:1995zx}.

Below we show that non-wave function vertices are supplanted by contributions from higher Fock states in light-front 
time-ordered perturbation theory (provided the interaction has light-cone time dependence). 
In essence contributions from non-wave function vertices are reducible and should 
only be used when the interaction is (or is approximately) instantaneous\footnote{We shall often refer to interactions and
vertices merely as instantaneous if they are independent of \emph{light-cone} time, or equivalently \emph{light-cone} energy.}. 
This constitutes a replacement theorem for non-wave function vertices which trivially extends to $(3+1)$ dimensions.

The organization of the paper is as follows. First in section \ref{LFCQM} we present the issue of non-wave function vertices and 
energy poles of the Bethe-Salpeter vertex focusing on the light-front constituent quark model as an example. Non-wave function 
vertices are required to express the form factor. We find the commonly used assumptions in quark models necessitate 
vertices not only without energy poles but without energy dependence. Next in section \ref{reduce}
we review the reduction of the Bethe-Salpeter equation presented in \cite{Sales:1999ec} focusing on the energy poles of the vertex.
We derive an interpretation of the reduction as a procedure for approximating the poles of the vertex. Additionally in this section 
we construct the gauge invariant current to be used with the reduced formalism. 
In section \ref{meat} the ladder model is presented and we compare the calculation of the form factor for the model using two 
different paths to the reduction. The comparison allows us to see when non-wave function vertices can be efficiently used. 
Lastly we summarize our findings in section \ref{sum}.

\section{Poles of the Bethe-Salpeter vertex} \label{LFCQM}
\begin{figure}
\begin{center}
\epsfig{file=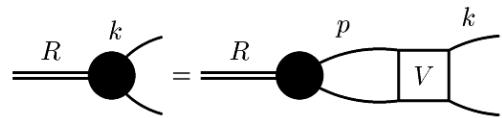}
\end{center}
\caption{Diagrammatic representation of the Bethe-Salpeter equation. The blob represents the vertex function $\Gamma$.}
\label{fBS}
\end{figure}

To introduce the reader to non-wave function vertices and instantaneous approximations, we focus on
light-front constituent quark models. We start by writing down the covariant equation for the meson vertex function $\Gamma$. It satisfies a simple 
Bethe-Salpeter equation \cite{Schwinger:1951ex} (see Figure \ref{fBS}):
\begin{equation} \label{BS}
\Gamma(k,R) =  i \int \frac{d^2p}{(2\pi)^2} V(k,p) \Psi_{BS}(p,R),
\end{equation}
in which we have defined the Bethe-Salpeter wave function $\Psi_{BS}$ as
\begin{equation}\label{psiBS}
\Psi_{BS}(k,R) = G(k,R) \Gamma(k,R),
\end{equation}
with the two-particle disconnected propagator $G(k,R) = d(k) d(R-k)$. For scalars of mass $m$, the renormalized, single-particle propagator $d$ 
has a Klein-Gordon form
\begin{equation}
d(k) = \frac{i}{(k^2 - m^2)[1+ (k^2 - m^2)f(k^2)]+ i \epsilon},
\end{equation}
where the residue is $i$ at the physical mass pole and the function $f(k^2)$ characterizes the renormalized, one-particle irreducible self-interactions. 
To simplify the comparison carried out in section \ref{meat}, we shall ignore $f(k^2)$. Above $V$ is the irreducible two-to-two scattering
kernel which we shall refer to inelegantly as the interaction potential. 

Now we imagine the initial conditions of our system are specified on the hypersurface $x^+= 0$.  
We define the plus and minus components of any vector by $x^\pm = \frac{1}{\sqrt{2}}(x^0 \pm x^3)$. Correlations between field operators 
evaluated at equal light-cone time turn up in hard processes for which this choice of initial surface is natural. In order to project out
the initial conditions (wave functions, \emph{etc}.) of our system, we must perform the integration over the Fourier conjugate to $x^+$, namely $k^-$.
For instance, our concern is with the light-front wave function defined as the projection of the covariant Bethe-Salpeter wave function onto $x^+ = 0$,
\begin{equation} \label{psilf}
\psi(x) = 2 R^+ x(1-x) \int \frac{dk^-}{2\pi} \Psi_{BS}(k,R),
\end{equation}
with $ x = \kplus / R^+$.

Looking at Eq. \eqref{psiBS}, in order to project the wave function exactly, we must know the analytic structure of the bound-state vertex function. If the vertex function $\Gamma(k,R)$ had no poles in $\kminus$, then our task would be simple: the light-front projection of $\Psi_{BS}$ would pick up contributions only from the poles of the propagator $G(k,R)$. Next we observe from Eq.~\eqref{BS}, that the $\kminus$ dependence of the interaction $V(k,p)$ must give rise to the $\kminus$ poles of the vertex function $\Gamma(k,R)$. Hence an instantaneous interaction gives rise to an instantaneous vertex and a simple light-front 
projection (see e.g.~\cite{Einhorn:1976uz}). 

On the other hand, constituent quark models often assume a less restrictive simplification of the analytic structure of the vertex 
(see e.g.~\cite{Jaus:zv}) in order to permit the light-cone projection. We shall show that this assumption
along with the presumed covariance of the quark model often implies instantaneous vertices. For any momentum $p$, let us denote
the on-shell energy $p^-_{\text{on}} = m^2 / 2 p^+$. The propagator $G(k,R)$ has two poles
\begin{equation} \label{valpoles}
 \begin{cases}
 k^-_{a} = k^-_{\text{on}} - \frac{\ie}{x}\\
 k^-_{b} = R^{-} + (k - R)^{-}_{\text{on}} - \frac{\ie}{x - 1}.
 \end{cases} 
\end{equation} 
Notice that although we work in $(1+1)$ dimensions, the results are trivial to extend to $(3+1)$ dimensions because the
imaginary parts of poles have precisely the same dependence on (only) the plus-momenta. This remark applies not only to this
section but to the rest of this work.

In constituent quark models $\Gamma$ is assumed to have no poles in the upper-half $k^-$-plane for $0<x<1$. Since the light-front wave function is 
proportional to $\theta[x(1-x)]$, in light of Eq.~\eqref{valpoles} we further require any poles of $\Gamma$ to lie in the upper-half plane for $x<0$ 
and in the lower-half plane for $x>1$. With these restrictions Eq.~\eqref{psilf} dictates the form of the constituent quark wave function
\begin{align} \label{g}
\psi(x) = \dw (x | M^2) \Gamma(k_{b},R) \theta[x(1-x)],
\end{align}
where we have defined the Weinberg propagator as
\begin{equation}
\dw (x|M^2) = \frac{1}{M^2 - \frac{m^2}{x(1-x)}}
\end{equation}
and used the abbreviation $\Gamma(k_{b},R)$ to denote evaluation at the pole $k^- = k^-_{b}$ appearing in Eq.~\eqref{valpoles}.

\begin{figure}
\begin{center}
	\epsfig{file=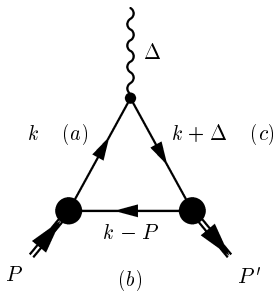,height=1.5in}
  	\caption{Covariant diagram for the electromagnetic form factor. Labels $a,b$ and $c$ denote
subscripts used for on-shell energy poles.}
\label{ftri}
\end{center}
\end{figure}

When we calculate the (elastic) electromagnetic form factor for these constituent quark models (see Figure \ref{ftri}), we are confronted with more poles. 
Let $\Delta^+ = - \zeta P^+ < 0$ be the plus-component of the momentum transfer between initial and final state mesons.  
The form factor is then
\begin{equation} \label{ff}
F(t) \propto \int \frac{(2 x - \zeta) \Gamma(k,P) \Gamma^*(k+\Delta,P+ \Delta) \; d^2 k}{[k^2 -m^2][(P-k)^2 - m^2][(k+\Delta)^2 -m^2]}
\end{equation}
The $k^-$-poles from the propagators are $k^-_{a}, k^-_{b}$ defined in \eqref{valpoles} and 
$k^-_{c} = - \Delta^- + (k+\Delta)^-_{\text{on}} - \frac{\ie}{x - \zeta}$.

Given this pole structure, the contributions to $F(t)$ are proportional to $\theta[x(1-x)]$. In the region $x<\zeta$, closing the contour in the 
upper-(lower-) half plane will enclose possible poles of the final (initial) vertex. 
As one can see from considering the region $x>\zeta$, the form factor  can be determined solely from $\Res(k^-_{b})$. 
In $(3+1)$ dimensions, where we are free to choose frames in which $\zeta = 0$, this is the only contribution to the form factor. 
But if $\zeta \neq 0$, the additional poles from vertices in the region $x < \zeta$ are required by Lorentz invariance. 

The authors \cite{Ji:2000fy} advocate no modification to the pole structure of Eq.~\eqref{ff} due to the vertices. Closing the contour in the lower-half
plane, they pick up the residue at $k^-_{a}$ without any contributions from poles of the initial-state vertex. Such poles cannot lie in the upper-half plane
for $x<\zeta$ since the form of $\psi$ would be both frame dependent and contrary to that of Eq.~\eqref{g} 
(the $(3+1)$-dimensional version of which is employed by \cite{Ji:2000fy}). 
Thus the authors are actually assuming there are no poles of the Bethe-Salpeter vertex \cite{Tiburzi:2001je}. 

Returning to the definition of the wave function \eqref{psilf} under the premise of a vertex devoid of poles, we now find $\Gamma(k_b,P) = 
\Gamma(k_a,P)$, which we shall call pole symmetry. This symmetry is essential for making contact with the 
Drell-Yan formula \cite{Drell:1970km}, since for $x>\zeta$ both 
initial- and final- state vertices may be expressed in terms of wave functions. When $x<\zeta$, however, the final-state vertex
becomes $\Gamma^*(k_a + \Delta, P + \Delta)$ which cannot be expressed in terms of $\psi^*(x^\prime)$ (where $x^\prime = \frac{x - \zeta}{1 - \zeta}$)
even with pole symmetry. Such a vertex is referred to as a non-wave function vertex. Understanding and dealing with such objects from 
the perspective of time-ordered perturbation theory is the primary goal of this paper. 

Because these constituent quark models are presumed covariant, Eq.~\eqref{ff} converges. This enables us to relate an initial-state
non-wave function vertex to the final-state non-wave function vertex encountered above via integrating around a circle at infinity. 
Equating the sum of residues $\Res (k^-_a) + \Res (k^-_b) + \Res (k^-_c) = 0$, we find
\begin{equation} \label{loop}
\frac{\Gamma^*(k_a + \Delta, P^\prime)}{\Gamma^*(k_b+\Delta,P^\prime)} (k^-_b - k^-_c)
- \frac{\Gamma(k_c,P)}{\Gamma(k_b,P)} (k^-_b - k^-_a) = k^-_a - k^-_c.
\end{equation}
This relation holds for the class of models for which the pole structure of Eq.~\eqref{ff} is not modified by the vertices. 
The ratio structure of the vertices does not allow
for a common factor $k^-_a - k^-_c$ in the two terms in Eq.~\eqref{loop}. The equality then depends on delicate cancellations
between initial- and final- state vertices, which in general are unrelated. The philosophy of constituent quark model 
phenomenology is to choose the form of $\psi$ and hence the form of $\Gamma$. Treating $\Gamma$ free, equality can only 
hold if both ratios are one. This yields the restriction
\begin{equation}
\frac{\partial}{\partial \Delta^-} \Gamma(k_c,P) = 0.
\end{equation}
But $\Delta^-$ only enters $\Gamma(k_c,P)$ through $k^- = k^-_c$. Hence $\Gamma$ is independent of $k^-$. In general, of course, 
the vertex $\Gamma$ not only has light-front energy dependence but poles as well. As we shall see below, the light-front reduction 
of the Bethe-Salpeter equation is a procedure for approximating the poles of the vertex function. Moreover when applied to current 
matrix elements, these poles generate higher Fock state contributions.

\section{Light-front reduction} \label{reduce}
In this section, we review the reduction scheme set up in \cite{Sales:1999ec} since their notation, while useful, is unfamiliar.
Additionally we resolve a peccadillo intrinsically related to non-wave function vertices. Next we 
give an intuitive picture for the reduction and finally construct the gauge invariant current for the calculation of form 
factors on the light-front. 

Above we have removed overall momentum-conserving delta functions, 
e.g. our propagator $G(k,R)$ is the momentum space version of $G(R)$ 
defined by $\langle R^\prime | G | R \rangle = (2 \pi)^2 \delta^{(2)}(R^\prime - R) G(R)$. 
In terms of these fully two-dimensional quantities, the Lippmann-Schwinger equation for the two-particle transition matrix $T$ appears as
\begin{equation} \label{LS}
T = V + V G T.
\end{equation}
A pole in the $T$-matrix (at some $R^2 = M^2$, say) corresponds to a two-particle bound state. Investigation of the pole's residue
gives the Bethe-Salpeter equation for the bound-state vertex $\Gamma$
\begin{equation} \label{Gamma}
\Gamma = V G \Gamma.
\end{equation}
The Bethe-Salpeter amplitude $\Psi$ is defined to be $G \Gamma$. 
Following \cite{Sales:1999ec}, we denote quantities able to be rendered in position or momentum space with 
bras and kets. We will employ this notation only for quantities that have been stripped of their overall 
momentum-conserving delta functions, for example $\Gamma(k,R) = \langle k | \Gamma_{R} \rangle $, 
where $R$ is used as a label for the bound state for which $R^2 = M^2$.

\subsection{The scheme}
To reduce the Lippmann-Schwinger equation to a light-front version, we must introduce an auxiliary Green's function
$\Gt$ in place of $G$ (as in \cite{Woloshyn:wm}). Thus we have
\begin{equation} \label{TW}
T = W + W \Gt T,
\end{equation}
provided that
\begin{equation} \label{W}
W = V + V (G - \Gt) W. 
\end{equation}
Taking residues of Eq.~\eqref{TW} gives us an alternate way to express the bound state vertex function 
\begin{equation} \label{regamma}
\Gamma = W \Gt \Gamma. 
\end{equation}

To choose a light front reduction, $\Gt$ must inherently be related to projection onto the initial surface $x^+ = 0$. For simplicity, we denote
the integration $\int \frac{d\kminus}{2\pi} \langle \kminus | \mathcal{O}(R) = \Big| \mathcal{O}(R)$. With this notation, 
we will always work in $(1+1)$-dimensional momentum space for which the only sensible matrix elements of $\Big| \mathcal{O}(R)$ are of the form 
$\langle \kplus | \;  \Big| \mathcal{O}(R) | \pminus, \pplus \rangle$. The operator $\mathcal{O}(R) \Big|$ is defined similarly. 
For a useful reduction scheme (one that preserves unitarity), 
we must have $\Big| G(R) \Big| = \Big| \Gt(R) \Big|$. 
%Bearing in mind an extra delta function in $G(R)$ relative to section \ref{LFCQM}, 
%we have already calculated $\Big| G(R) \Big|$ in Eq. \eqref{g} above, thus
%\begin{align}
%\langle \kplus | g(R) | \pplus \rangle & \equiv \langle \kplus | \; \Big| \Gt(R) \Big| \; | \pplus \rangle \notag \\
%& = \delta (\kplus - \pplus) \frac{\pi i \; \theta[x(1-x)]}{R^+ x(1-x)} \dw(x | R^2).
%\end{align} 
The simplest choice of $\Gt$ that results in time-ordered perturbation theory requires
\begin{equation} \label{gt}
\Gt(R) = G(R) \Big| g^{-1}(R) \; \Big| G(R),
\end{equation} 
where the reduced disconnected propagator $g(R)$ is defined by the matrix elements
\begin{equation} \label{lilg}
\langle x R^+ | \; \Big| G(R) \Big| \; |  y R^+ \rangle =  
\langle x R^+ | g(R) | y R^+ \rangle,
\end{equation}
explicitly this forces
\begin{multline}
\langle x R^+ | g(R) | y R^+ \rangle =  2\pi \delta(xR^+ - yR^+) \\
\times \theta[x(1-x)] \frac{2 \pi i}{2 R^+ x(1-x)} \dw(x|R^2).
\end{multline}
The inverse propagator $g^{-1}(R)$ can then be constructed, bearing in mind $g^{-1}(R)$ only exits in the subspace
where $g(R)$ is non-zero. Explicitly
\begin{multline} \label{lilginv}
\langle x R^+ | g^{-1}(R) | y R^+ \rangle =  2\pi \delta(xR^+ - yR^+) \\
\times \theta[x(1-x)] \frac{2 R^+ x(1-x)}{2 \pi i} \dw^{-1}(x|R^2).
\end{multline}
This forces 
\begin{equation}
\langle x R^+ | g^{-1}(R) g(R) | y R^+ \rangle =  2\pi \delta(xR^+ - yR^+) \theta[x(1-x)],
\end{equation}
which is unity restricted to the subspace where the operators $g(R)$ and $g^{-1}(R)$ are defined.
We are more careful about this point than the authors \cite{Sales:1999ec} 
%treat $g^{-1}(R)$ inconsistently. It appears $\propto \theta[x(1-x)]$
%when auxiliary quantities are calculated and then unthinkably $\propto 1/\theta[x(1-x)]$ when 
%reduced kernels are calculated. We belabor this point 
since the consequences of Eq.~\eqref{lilginv} are essential
for dealing with instantaneous interactions. Notice $\Gt(R)$ defined in Eq.~\eqref{gt} 
is non-zero only for plus-momentum fractions between zero and one. 

The reduced transition matrix $t(R)$ is
\begin{equation} \label{tred}
t(R) = g^{-1}(R) \Big| \Gt(R) T(R) \Gt(R) \Big| g^{-1}(R) 
\end{equation}
Taking the residue of Eq.~\eqref{tred} at $R^2 = M^2$, gives a homogeneous equation for the reduced vertex function $\gamma$
\begin{equation} \label{gammaredu}
| \gamma_{R} \rangle = w(R) g(R) | \gamma_{R} \rangle,
\end{equation}
where the reduced auxiliary kernel is
\begin{equation} \label{wred}
w(R) = g^{-1}(R) \Big| G(R) W(R) G(R) \Big| g^{-1}(R).
\end{equation}
Given this structure, the reduced kernel $w(x,y|R^2) \equiv \langle x R^+| w(R) | y R^+ \rangle$
will always be $\propto \theta[x(1-x)] \theta[y(1-y)]$. Moreover the reduced vertex $\gamma(x|M^2) 
\equiv \langle x R^+ | \gamma_R \rangle \propto \theta[x(1-x)]$ as a result of Eq.~\eqref{gammaredu}.

From \eqref{gammaredu} we can define the light-front wave function $|\psi_{R}\rangle \equiv g(R) |\gamma_{R}\rangle$, notice this too 
restricts the momentum fraction $x$: $\psi(x) \propto \theta[x(1-x)]$. 
By iterating the Lippmann-Schwinger equation for $T$ twice, it is possible to relate $T$ to $t$ and thereby construct $T$ given $t$, which is 
clearly not possible from the definition \eqref{tred}. Taking the residue of this relation between $T$ and $t$ yields the  
reduced-to-covariant conversion between bound-state vertex functions, namely
\begin{equation} \label{convert}
| \Gamma_{R} \rangle = W(R) G(R) \Big| \;| \gamma_{R} \rangle.
\end{equation}
Finally, we can manipulate the covariant Bethe-Salpeter amplitude into the form
\begin{equation} \label{324}
| \Psi_{R} \rangle = \Bigg( 1 + \Big(G(R) - \Gt(R)\Big)W(R) \Bigg) G(R) \Big| \; |\gamma_{R} \rangle,
\end{equation}
which justifies the interpretation of $|\psi_{R}\rangle$ as the light-front wave function since
$\Big| \; |\Psi_{R}\rangle = 
%\Bigg(  g(R) + \underbrace{\Big( \Big| G(R) - \Big| \Gt(R)\Big)}_{0}W(R) G(R) \Big|  \Bigg) |\gamma_{R} \rangle = 
|\psi_{R}\rangle$.

While all light-front reduction schemes when summed to all orders yield the 
$x^+ = 0$ projection of the Bethe-Salpeter equation,  the choice of $\Gt(R)$
in Eq.~\eqref{gt} generates a kernel calculated in light-front time-ordered perturbation theory.
Lastly the normalization of the covariant and reduced wave function is discussed in \cite{future}.

\subsection{An interpretation for the reduction} \label{intuit}
The heart of our intuition about the light cone lies in integrating out the minus-momentum dependence of the 
covariant wave function.  So we merely cast the formal reduction in a way which highlights the contributions 
from poles of the vertex function.

Utilizing Eqs. \eqref{regamma} and \eqref{convert},
% while omitting total four-momentum labels since they 
%are all identical, 
we can show
\begin{equation} \label{key}
\Gt (R) | \Gamma_R \rangle = G(R) \Big| \; |\gamma_R \rangle.
\end{equation}  
Thus the appearance of $\Gt(R) | \Gamma_R \rangle$ has the form of an instantaneous approximation since Eq.~\eqref{key} shows
that it has no minus-momentum poles besides those of the propagator $G(R)$. 

This instantaneous approximation appears in determining the light-front wave function. Using Eqs.~\eqref{W} and \eqref{324}, we
have
\begin{multline}
| \psi_R \rangle %= \Big| \; | \Psi_R \rangle  
= \Big| G(R) V(R) \\
\times \sum_{j = 0}^{\infty} \Big[ \Big( G(R) - \Gt(R) \Big) V(R) \Big]^j G(R) \Big| \;  |\gamma_R \rangle.
\end{multline}
From truncating the series in $G(R) - \Gt(R)$ at some $j = n - 1$ and using a consistent approximation to Eq.~\eqref{convert}, 
we are led to the approximate solution 
\begin{equation} \label{interpretation}
|\psi^{(n)}\rangle  = \Big| (GV)^n \Gt | \Gamma \rangle,
\end{equation}
after having used Eq.~\eqref{key}.
Thus at any order $n$ in the formal reduction scheme, 
we have iterated the covariant Bethe-Salpeter equation $n$-times and subsequently made an instantaneous approximation via Eq.~\eqref{key}.
Retaining the minus-momentum dependence in $V$ to $n^{\text{th}}$ order allows for an $n^{\text{th}}$-order approximation to the vertex function's poles. 

\subsection{Current in the reduced formalism}\label{current}
Here we extend the formalism presented so far to include current matrix elements between bound states. We do so in a gauge invariant 
fashion following \cite{Gross:bu}. Our notation, however, is more in line with the elegant method of gauging equations presented in 
\cite{Kvinikhidze:1998xn}. This latter general method extends to bound systems of more than two particles. 

\begin{figure}
\begin{center}
\epsfig{file=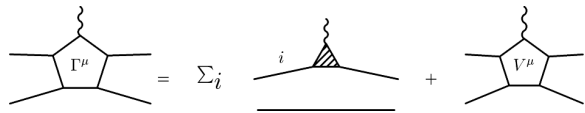,width=3in}
\end{center}
\caption{Graphical depiction of the irreducible five-point function $\Gamma^\mu$ as 
sum of impulse terms and a gauged interaction. By construction, 
$\Gamma^\mu$ is gauge invariant.}
\label{fgamma}
\end{figure}

Consider first the full four-point function $G^{(4)}$ defined by
\begin{equation}\label{g4}
G^{(4)} = G + G T G. 
\end{equation}
For later use, it is important to note that the residue of $G^{(4)}(R)$ at the bound state pole $R^2 = M^2$ is $- i |\Psi_{R}\rangle\langle \Psi_{R}|$. 
Using the Lippmann-Schwinger equation for $T$, we can show the four-point function satisfies 
\begin{equation}\label{LSg4}
G^{(4)} = G + G V G^{(4)}. 
\end{equation}

To discuss electromagnetic current matrix elements, we will need the three-point 
function $d_{i}^\mu$ where the label $i$ denotes particle number. We define an irreducible
three-point function $\Gamma_{i}^\mu$ in the obvious way
\begin{equation}
d_{i}^\mu = d_{i} \Gamma_{i}^\mu d_{i}.
\end{equation}
Now we need to relate the one-particle electromagnetic vertex function to the $T$ matrix. Let $j^\mu$ denote the electromagnetic coupling to the constituent
particles (since our particles are scalars $j^\mu = \overset{\leftrightarrow}\partial{}^\mu$). 
Since the electromagnetic three-point function $\Gamma_{i}^\mu$ is irreducible, we have
\begin{equation}
\Gamma_{i}^\mu  = G^{-1} G^{(4)} j^\mu
%\overset{\leftrightarrow}\partial{}^\mu 
\end{equation}
 and by using the definition of $G^{(4)}$ Eq.~\ref{g4}, we have the desired relation
\begin{equation}
\Gamma_{i}^{\mu} = %\overset{\leftrightarrow}\partial{}^\mu  
j^\mu + T G j^\mu
%\overset{\leftrightarrow}\partial{}^\mu.
\end{equation}
Notice the right hand side lacks the particle label $i$. In the first term, the bare coupling acts on the $i$th particle while in the second term the bare coupling does not act on the $i$th particle.
For this reason we have dropped the label which will always be clear from context.

In considering two propagating particles' interaction with a photon, the above definitions lead us to the impulse approximation
to the current
\begin{equation}
\Gamma_{0}^\mu = \Gamma_{1}^\mu d_{2}^{-1} + d_{1}^{-1} \Gamma_{2}^\mu.
\end{equation}
Additionally the photon could couple to interacting particles. Define a gauged interaction $V^\mu$ topologically
by attaching a photon to the kernel in all possible places. This leads us to the irreducible electromagnetic vertex $\Gamma^\mu$ defined as (see Figure \ref{fgamma})
\begin{equation} \label{emvertex}
\Gamma^\mu = \Gamma_{0}^\mu + V^\mu,
\end{equation}
which is gauge invariant by construction.

\begin{figure}
	\begin{center}
	\epsfig{file=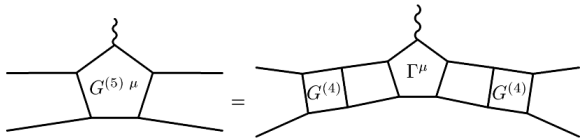,width=3in}
		\end{center}
	\caption{Graphical depiction of the five-point function $G^{(5) \; \mu}$. The 
	irreducible five-point function is the gauge invariant $\Gamma^\mu$.}
	\label{f5pt}
\end{figure}

Lastly to calculate matrix elements of the current between bound states it is useful to define a reducible five-point function
(see Figure \ref{f5pt})
\begin{equation} \label{5alive}
G^{(5) \; \mu} = G^{(4)} \Gamma^\mu G^{(4)}.
\end{equation}
Having laid out the necessary facts about electromagnetic vertex functions and gauge invariant currents, we can now specialize to their matrix elements 
between bound states by taking appropriate residues of Eq.~\eqref{5alive}. The form factor is then
\begin{equation} \label{me}
- i (P^{\prime\mu} + P^\mu) F(t) = \langle \Psi_{P^\prime} | \Gamma^\mu(-\Delta) | \Psi_{P} \rangle, 
\end{equation}
where $P^{\prime \mu} = P^\mu + \Delta^\mu$ and $t = \Delta^2$. 

\section{Two models} \label{meat}

\subsection{Wave functions}
\begin{figure}
\begin{center}
\epsfig{file=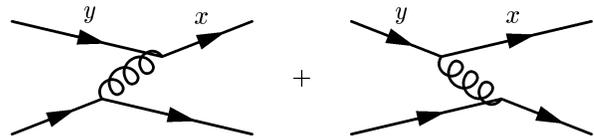}
\end{center}
\caption{Diagrammatic representation of the one-boson exchange potential $w(x,y|M^2)$ appearing in Eq. \eqref{OBEP}.}
\label{fOBEP}
\end{figure}

To say anything less than general, we must know the minus-momentum dependence of the interaction. We therefore adopt a weakly coupled, 
one-boson exchange model for $V$ (the so-called ladder approximation). Supposing the boson mass is $\mu$ and the coupling constant $g$, we have
\begin{equation} \label{ladder}
V(k,p) =  \frac{- g^2}{(p-k)^2 - \mu^2 + \ie}  
\end{equation}
where the energy pole (with respect to $p$) of the interaction is 
\begin{equation} \label{pv}
p^-_{v} = \kminus + \frac{\mu^2}{2 (\pplus - \kplus)} -\frac{\ie}{2(\pplus - \kplus)}.
\end{equation}
This interaction is non-local in space-time and hence does not have an instantaneous piece. The reduced kernel Eq.~\eqref{wred} 
is consequently made up of retarded terms (i.e.~dependent on the eigenvalue $M^2$) where higher order in $G - \Gt$ 
means more particles propagating at a given instant of light-cone time (see, e.g.~\cite{Sales:1999ec}). 
%As is known \cite{Sales:1999ec}, 
The leading-order equation for $\psi$ from Eq.~\eqref{gammaredu} 
is the Weinberg equation \cite{Weinberg:1966jm}
\begin{equation} \label{OBE}
\psi(x) = - \dw(x|M^2) \int_0^1 \frac{w(x,y|M^2) \psi(y)}{2 (2\pi) y(1-y)} dy.
\end{equation}
with the time-ordered one-boson exchange potential (see Figure \ref{fOBEP}) calculated to leading order from 
Eq.~\eqref{wred}
\begin{multline} \label{OBEP}
w(x,y|M^2) = \frac{- g^2}{x - y} \theta[x(1-x)] \theta[y(1-y)] \\ 
\times \Big[ \theta(x-y) D(x,y|M^2) - \{x \leftrightarrow y\}  \Big],
\end{multline}
where
\begin{equation}
D^{-1}(x,y|M^2) = M^2 - \frac{m^2}{y} - \frac{\mu^2}{x - y} - \frac{m^2}{1-x}.
\end{equation}
We can obtain Eq.~\eqref{OBE} most simply by iterating the Bethe-Salpeter equation once (see section \ref{intuit}) 
and then projecting onto the light cone. 
%We have left off a factor of $\theta[x(1-x)]$ produced by the contour 
%integration and will treat this factor as implicit on any wave function.

In the limit $\mu^2 \gg m^2, M^2$, the interaction becomes approximately 
instantaneous, which suggests we separate out an instantaneous piece $V_o$:
\begin{equation} \label{instladder}
V(k,p) = V_o + [V(k,p) - V_o],
\end{equation}
where
\begin{equation} \label{Vo}
V_o = V_o(x,y) = \frac{- g^2 \theta(x - y)}{E^2(x - y) - \mu^2} + \frac{-g^2 \theta(y-x) }{E^2(y - x) - \mu^2},
\end{equation}
with $E$ as a parameter to be chosen. Of course other choices of $V_o$ are possible. 
%\footnote{Of course, other choices are possible, 
%e.g.~$V_o = \frac{-g^2}{E^2 (x - y) - \mu^2}$ where $E^2$ is a parameter to be chosen. The optimal choice
%of $V_o$ is not under investigation here.} $V_o = \frac{g^2}{\mu^2}$. In this case we expand Eq.~\eqref{W} simultaneously in 
%$G - \Gt$ as well as in $V - V_o$. 
We choose the above form of $V_o$ for two reasons. First is the form of the instantaneous approximation wave function $\phi(x)$. 
%\subsection{Wave functions}
%In order to have a useful comparison of instantaneous and non-instantaneous contributions, we will choose $g^2$ to be suitably small
%so that Eq.~\eqref{W} can be well approximated to first order in $G - \Gt$. This enables systematic organization of our observations.
When we write the potential as Eq.~\eqref{instladder}, we expand Eq.~\eqref{W} to first order in $g^2$ as
\begin{equation} \label{instW}
W = V_o + (V - V_o ) + (V - V_o)\Big( G - \Gt \Big) (V - V_o). 
\end{equation}
To zeroth order in $G - \Gt$ and $V - V_o$, the equation for $\phi$ is
\begin{equation} \label{nOBE}
\phi(x) = - \dw(x|M^2) \int_0^1 \frac{ w_o(x,y) \phi(y)}{2(2\pi) y(1-y)} dy, 
\end{equation} 
where by Eq.~\eqref{wred}, the reduced instantaneous potential is merely $w_o(x,y) = \theta[x(1-x)] \theta[y(1-y)] V_o(x,y)$.
%%%%%and hence $\phi$ is just proportional to a propagator $\dw$. This exactly soluble light-cone model has been considered before
%in \cite{Sawicki:hs}. 
In the instantaneous limit, $\mu^2 \gg m^2, M^2$ solutions to Eqs.~\eqref{OBE} and \eqref{nOBE} 
coincide---both wave functions approach $\sim \dw(x,4m^2)$. Secondly, we 
preserve the contact interaction limit by excluding from $V_o$ the factor $\theta[x(1-x)]\theta[y(1-y)]$.
That is, when $\{\mu^2,g^2\} \to \infty$ with $g^2/\mu^2$ fixed (or equivalently $E \to 0$) we have the contact interaction\footnote{This 
exactly soluble $(1+1)$-dimensional light-cone model, in which $\psi(x) \propto \dw(x|M^2)$, 
has been considered earlier in \cite{Sawicki:hs}.}
$V_o \to g^2 /\mu^2$, which rightly knows nothing about the momentum fractions $x$ and $y$.    
%Moreover our concern is to investigate the instantaneous limit for form factors not wave functions. Thus to put the wave functions
%on equal footing we shall solve for $\phi$ additionally to first order in $V - V_o$, i.e.~$\phi(x) = \psi(x)$. 

We shall now investigate contributions to the %wave functions and 
form factors for each of the 
models Eqs.~\eqref{ladder} and \eqref{Vo}. 	

\subsection{Form factors}
From Eq.~\eqref{me} we can calculate the form factors for each of the models \eqref{ladder} and \eqref{Vo} in the reduced formalism. 
%Let $\Delta^\mu$ denote the momentum transfer. 
Working in perturbation theory, we separate out contributions up to first order
by using Eq.~\eqref{324} to first order in $G - \Gt$ and $\Gamma^\mu$ \eqref{emvertex} in the first Born approximation. The 
matrix element $J^\mu = \langle \Psi_{P^\prime} | \Gamma^\mu(-\Delta) | \Psi_{P} \rangle$ then appears for a model with some kernel $V$

\begin{multline}
J^\mu  \approx  \; \langle \gamma_{P^\prime} | \; \Big| G(P^\prime) \Big( 1 + V(P^\prime) ( G(P^\prime) - \Gt(P^\prime)) \Big) 
\\ \times \Big( \overset{\leftrightarrow}\partial{}^\mu(-\Delta)  d_{2}^{-1} + V(-\Delta) G(-\Delta) 
\overset{\leftrightarrow}\partial{}^\mu(-\Delta) d_{2}^{-1} \Big) 
\\ \times \Big( 1 + (G(P) - \Gt(P)) V(P)  \Big) G(P) \Big| \; | \gamma_{P} \rangle  \\
	= \Big( J^\mu_{\text{LO}} + \delta J^\mu_{i} +  \delta J^\mu_{f} + \delta J^\mu_{\gamma} \Big) + \mathcal{O}[V^2],
\end{multline}
%(factors of $2$ stem from identical constituents) 
with the leading-order result
\begin{equation} \label{LO}
J^\mu_{\text{LO}} = \langle \gamma_{P^\prime} | \; \Big| G(P^\prime) \overset{\leftrightarrow}\partial{}^\mu (-\Delta) 
d_{2}^{-1} G(P) \Big| \; | \gamma_{P} \rangle.
\end{equation}
The first-order terms are
\begin{multline}    
		\delta J^\mu_{i} =  \langle \gamma_{P^\prime} | \; \Big| G(P^\prime) \overset{\leftrightarrow}\partial{}^\mu (-\Delta) 
		d_{2}^{-1} 
		\\ \times \Big(G(P) - \Gt(P) \Big) V(P) G(P) \Big| \; |\gamma_{P} \rangle \notag 
\end{multline} 
\begin{multline}  
		\delta J^\mu_{f} = \langle \gamma_{P^\prime} | \; \Big| G(P^\prime) V(P^\prime) \Big(G(P^\prime) - \Gt(P^\prime) \Big) 
		\\ \times \overset{\leftrightarrow}\partial{}^\mu (-\Delta) d_{2}^{-1} G(P) \Big| \; | \gamma_{P} \rangle \notag 
\end{multline} 
\begin{multline}  
		\delta J^\mu_{\gamma}  = \langle \gamma_{P^\prime} | \; \Big| G(P^\prime) \Big(V(-\Delta) G(-\Delta) 
		\overset{\leftrightarrow}\partial{}^\mu (-\Delta) \Big) \\
		\times d_{2}^{-1} G(P) \Big| \; | \gamma_{P} \rangle. \label{NLO}
\end{multline} 
The labels indicate the intuition behind the reduction scheme (seen in section \ref{intuit}):
the term $\delta J^\mu_{f}$ arises from one iteration of the covariant Bethe-Salpeter equation for the
final-state vertex followed by an instantaneous approximation \eqref{key}, $\delta J^\mu_{i}$ arises in 
the same way from the initial state and $\delta J^\mu_\gamma$ comes from one iteration of the Lippmann-Schwinger
equation \eqref{LS}. 

Because the leading-order expression \eqref{LO} is independent of the kernel $V$, the result will have the same form for both models. 
Using the effective resolution of unity, the above expression converts into
\begin{multline}
J^\mu_{\text{LO}} = \int \frac{d^2 p}{(2\pi)^2} \; \frac{d^2k}{(2\pi)^2} \langle \gamma_{P^\prime} | \pplus\rangle 
\\ \times \langle p | G(P^\prime)\overset{\leftrightarrow}\partial{}^+ (-\Delta) d_{2}^{-1} G(P) | k \rangle \; \langle \kplus | \gamma_{P} \rangle.
\end{multline} 
Bearing in mind the delta function present in $G(R)$, we have the factor
\begin{equation}
\langle p | \overset{\leftrightarrow}\partial{}^+ (-\Delta)| k \rangle = -i (2 \kplus + \Delta^+) (2\pi)^2 \delta^{2}(p - k - \Delta) .  
\end{equation}
Now define $ x = \kplus/P^+$ and $\Delta^+ = - \zeta P^+$ as above and denote the reduced vertex 
$\langle \kplus| \gamma_{R} \rangle = \gamma(x|M^2)$. The leading-order contribution is then
\begin{multline} \label{formLO}
J^+_{\text{LO}} = \int \frac{d^2 k}{(2 \pi)^2}  \gamma^*(x^\prime|M^2)  d(k + \Delta) 
\\ \times \big( 2 x - \zeta \big) d(k) d(P-k) \gamma(x|M^2),
\end{multline}
where $x^\prime = \frac{x - \zeta}{1-\zeta}$. Eq.~\eqref{formLO} is quite similar to \eqref{ff}. For $x>\zeta$ evaluation
is straightforward and leads to 
\begin{equation}\label{FFLO}
F_{LO}(t) = \frac{\theta(x - \zeta)}{1 - \zeta/2} \int \frac{dx}{2 (2\pi)} \frac{2 x - \zeta}{x (1-x) x^\prime} \psi^*(x^\prime) \psi(x),
\end{equation}
for the non-instantaneous case. For the instantaneous case $F_{LO}(t)$, replace $\psi$ with $\phi$.

\begin{figure}
\begin{center}
\epsfig{file=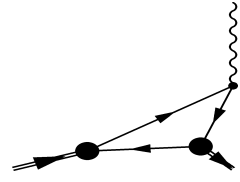,width=1.2in,height=0.9in}
\caption{The $Z$-graph confronting evaluation of the electromagnetic form factor}
\label{fZ}
\end{center}
\end{figure}

On the other hand, we know from section \ref{LFCQM} the region $x<\zeta$ contains a non-wave function vertex
(see Figure \ref{fZ}).
However, evaluation of Eq.~\eqref{formLO} with reduced vertices is subtly different than in section \ref{LFCQM}. 
Quite simply, the term $\gamma^*(x^\prime|M^2) = 0$ by virtue of Eq.~\eqref{gammaredu} because $x^\prime < 0$. Thus there
is no contribution at leading order for $x < \zeta$.

The first-order terms depend explicitly on the interaction and will hence be considerably different for each of the models. 
We consider each model separately.

\subsubsection{Non-instantaneous case}

\begin{figure}
\begin{center}
\epsfig{file=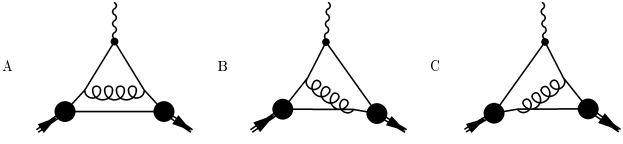,width=3.25in}
\caption{Diagrams which contribute to the form factor to first order in $G - \Gt$ for the non-instantaneous case (for $x >\zeta$).}
\label{ftri2}
\end{center}
\end{figure}

\begin{figure}
\begin{center}
\epsfig{file=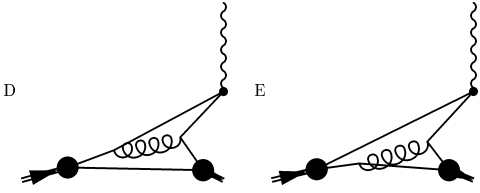,width=2.5in}
\caption{The remaining diagrams (characterized by $x < \zeta$) for the electromagnetic form factor at first order in $G - \Gt$ for the 
non-instantaneous case.}
\label{fZZZ}
\end{center}
\end{figure}

Evaluating contributions at first order for the non-instantaneous interaction Eq.~\eqref{ladder} is complicated
by the presence of poles in the interaction (\emph{cf} Eq.~\eqref{pv}). First we evaluate the first Born term
$\delta J^\mu_{\gamma}$ in Eq.~\eqref{NLO}. After careful evaluation of the two minus-momentum integrals,
%keeping terms only of order $g^2$
we have the contribution to $\delta J^+_\gamma$ for $x>\zeta$
\begin{multline}
\delta J^+_{A} = \int \frac{\theta(x - \zeta) \; dx dy \; (2 x - \zeta)}{16\pi^2 x x^\prime y(1-y) y^\prime} \psi^*(y^\prime) \\ \times 
D(y^\prime,x^\prime|M^2) \frac{g^2 \theta(y - x)}{y - x} D(y,x|M^2) \psi(y),
\end{multline}
where $y^\prime = \frac{y - \zeta}{1 - \zeta}$. This contribution corresponds to diagram $A$ in Figure \ref{ftri2}. 
Additionally using $x^{\prime\prime} = x / \zeta$, we have for $x<\zeta$
\begin{multline}
\delta J^+_D = \int \frac{\theta(\zeta -x) \; dx dy \; (2 x - \zeta)/\zeta}{16\pi^2  y(1-y) y^\prime x^{\prime\prime} (1 - x^{\prime\prime})}
\psi^*(y^\prime) \\ \times \dw(x^{\prime\prime}|t) \frac{g^2 \theta(y - x)}{y - x} D(y,x|M^2)  \psi(y),
\end{multline}
which corresponds to diagram $D$ in Figure \ref{fZZZ}.

The initial-state iteration term $\delta J^\mu_{i}$ in Eq.~\eqref{NLO} is complicated by the subtraction of the two-particle
reducible contribution
\begin{multline} \label{isreduce}
- \langle \gamma_{P^\prime} | \; \Big| G(P^\prime) \overset{\leftrightarrow}\partial{}^\mu (-\Delta) d_{2}^{-1} \Gt(P) V(P) G(P) 
\Big| \; |\gamma_{P} \rangle 
\\ = - \langle \gamma_{P^\prime} | \; \Big| G(P^\prime) \overset{\leftrightarrow}\partial{}^\mu (-\Delta) 
d_{2}^{-1} G(P) \Big| \; |\gamma_{P} \rangle. 
\end{multline}
Thus this term merely removes contributions which can be reduced into the initial-state wave function. 
Evaluation of the two minus-momentum integrals 
%at order $g^2$ 
yields a contribution for $x>\zeta$ to $\delta J^+_{i}$
\begin{multline}
\delta J^+_{B} =  \int \frac{\theta(x-\zeta) \; dx dy \; (2 x - \zeta)}{16\pi^2 x x^\prime (1-x^\prime) y (1-y)}
\psi^*(x^\prime) \\ \times  D(y^\prime,x^\prime|M^2) \frac{g^2 \theta(y - x)}{y - x} D(y,x|M^2) \psi(y),
\end{multline}
which corresponds to diagram $B$ in Figure \ref{ftri2}. On the other hand,  for 
$x<\zeta$, we have the non-wave function vertex $\gamma^*(x^\prime|M^2)$ for the final state, which vanishes.
Similarly the $\Gt(P)$ term vanishes. Thus there is no contribution to $\delta J^\mu_{i}$ for $x<\zeta$. 
%But to remain at order $g^2$, the contribution from $V(P)$ must be 
%reduced into the initial state. This contribution is then identical to Eq.~\eqref{formLO} above for $x<\zeta$ and is then removed
%by $\Gt$ in Eq.~\eqref{isreduce}.\footnote{We can consistently avoid this complication by treating 
%$\gamma(x^\prime|M^2) \propto \theta[(x^\prime(1-x^\prime)]$
%for which the term for $x<\zeta$ and its subtraction $\Gt$ are both zero.} 

Finally there is the final-state iteration term $\delta J^\mu_f$ in Eq.~\eqref{NLO}. There are only two types of contributions. For
$x>\zeta$, that which can be reduced in to the final-state wave function is subtracted by $\Gt$. 
%When $x<\zeta$, $\Gt$ is zero in Eq.\eqref{NLO}.
The remaining term is:
\begin{multline}
\delta J^+_{C} =  \int \frac{\theta(x - \zeta) \;dx dy^\prime (2 x - \zeta)}{16\pi^2 x(1-x) x^\prime y^\prime (1- y^\prime)} \psi^*(y^\prime)
\\ \times D(y^\prime,x^\prime|M^2) \frac{g^2 \theta(y - x)}{y - x} D(y,x| M^2) \psi(x),
\end{multline}
where $y = \zeta + (1-\zeta) y^\prime$. This corresponds to diagram $C$ in Figure \ref{ftri2}.  For $x<\zeta$, 
the subtraction term vanishes since $x^\prime < 0$ for which $\Gt(P^\prime) = 0$. 
The remaining term in $\delta J^+_{f}$ gives a contribution
\begin{multline}
\delta J^+_{E} =  \int \frac{\theta(\zeta - x) \;dx dy^\prime  (2 x - \zeta)/\zeta}{ 16\pi^2 (1-x) x^{\prime\prime} (1 - x^{\prime\prime})
y^\prime (1-y^\prime)} \psi^*(y^\prime) \\ \times \dw(x^{\prime\prime}|t)  \frac{g^2 \theta(y - x)}{y - x} 
D(y,x| M^2) \psi(x),
\end{multline}
which corresponds to diagram $E$ in Figure \ref{fZZZ}. To summarize, the non-valence correction to the form factor in the non-instantaneous
case is
\begin{equation} \label{NVNI}
\delta F_{NI} =  \frac{1}{1-\zeta/2} \Big[ \delta J^+_A + \delta J^+_B + \delta J^+_C + \delta J^+_D + \delta J^+_E \Big],
\end{equation}
and there are no non-wave function terms.

\subsubsection{Instantaneous case}
\begin{figure}
\begin{center}
\epsfig{file=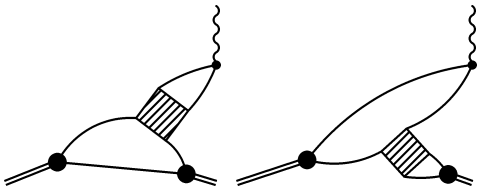,width=2.5in}
\caption{Diagrams with crossed interactions necessary to calculate the form factor in the region $x < \zeta$ for the instantaneous case.}
\label{fcross}
\end{center}
\end{figure}

The case of an instantaneous interaction is quite different due to the absence of light-front energy poles in $V_o$. 
Note we are working with Eq.~\eqref{instW} to zeroth order in $G - \Gt$ and $V - V_o$. 
%The conversion of the light-front wave function to the covariant wave function simplifies because there is no 
%minus-momentum dependence in $V_o$. In this case, we have a factor of $G(R) \Big| - \Gt(R) \Big| = 0$ in 
%Eq.~\eqref{324} which reduces to
%\begin{equation}
%|\Psi_R\rangle = G(R) \Big| \; | \gamma_R \rangle.
%\end{equation}
%Consequently contributions to the form factor $\delta J^\mu_{i}$ and $\delta J^\mu_f$ vanish. 
The first term in Eq.~\eqref{NLO} we consider, is the Born term $\delta J^\mu_\gamma$. 
The pole structure leads only to a contribution for $x<\zeta$:
%To order $g^2$ we have the contribution to $\delta J^+_\gamma$
\begin{multline} \label{btinst}
\delta J^+_1 = - \int \frac{\theta(\zeta - x) \theta(y - x) \; dx dy \; (2 x - \zeta)/ \zeta}{16\pi^2 y (1-y) y^\prime x^{\prime\prime}
(1- x^{\prime\prime})} 
\\ \times \phi^*(y^\prime) \dw(x^{\prime\prime}|t) V_o(x^{\prime \prime},y^{\prime \prime}) \phi(y),
\end{multline}
and is depicted by the first diagram in Figure \ref{fcross}. The interaction along the way to the photon vertex is crossed and
represents pair production off the quark line ($x^{\prime \prime} > 1$).

The next term $\delta J^\mu_{i}$ simplifies considerably due to the absence of light-cone time dependence in $V_o$.
As above, the final state vertex restricts $\zeta < x < 1$. But then we are confronted with a factor
$ \langle k | \Big(G(P) - \Gt(P)\Big) \Big| = 0$ since $x > 0$. Thus $\delta J^\mu_i = 0$. 
%While $\delta J^\mu_i = \delta J^\mu_f = 0$, it would seem we have exhausted 
%all first-order corrections to the form factor; however, there 
%is a subtlety. To double check, we iterate the Bethe-Salpeter equation for the initial and final states. For the initial state, 
%the iteration
%merely reduces back into the vertex by way of the bound-state equation \eqref{nOBE}. For the final state, the iteration reduces into the vertex for 
%$x>\zeta$, while for $x<\zeta$ we have Eq.~\eqref{cross} for the final-state vertex. The crucial difference from the non-instantaneous case 
%is that we obtain \eqref{cross} from covariant iteration (alternatively, since $V_o$ is instantaneous we can view \eqref{cross} as a covariant iteration). 

The last term we must consider is $\delta J^\mu_f$, in which we have the factor
$\Big| \Big( G(P^\prime) - \Gt(P^\prime) \Big) | k + \Delta \rangle$. This is zero for $x - \zeta > 0$, else $\Gt(P^\prime) = 0$ by 
virtue of Eq.~\eqref{gt} and Eq.~\eqref{lilginv}. Thus we only have a contribution for $x<\zeta$ which is from 
$\Big| G(P^\prime) | k+\Delta \rangle$. The expression for this contribution is
\begin{multline} \label{fsiinst}
\delta J^+_2 = - \int \frac{\theta(\zeta - x) \theta(y - x) dx d\yp (2 x - \zeta) / \zeta}{16 \pi^2 (1-x) \xpp (1-\xpp) \yp (1-\yp)}\\
\times \phi^*(\yp) w_o(y^\prime,x^\prime) \dw(\xpp|t) \phi(x),
\end{multline}
and is depicted on the right in Figure \ref{fcross}. The interaction again is crossed ($x^\prime <0$) and represents pair production.
With Eqs.~\eqref{btinst} and \eqref{fsiinst}, we have both bare-coupling pieces of the 
full Born series for the photon vertex (further terms in the series, which result from higher order terms in the expansion of $W$, 
add interaction blocks to each diagram on the quark-antiquark pair's 
path to annihilation). In this way we recover the Green's function from summing the Born series \cite{Einhorn:1976uz,Tiburzi:2001je}. 
%(for which the leading term Eq.~\eqref{fsiinst} is essential). 
Notice also the above form of $\delta J^+_2$ is what one would obtain from extending the definition of $\gamma$ as a non-wave
function vertex \cite{Einhorn:1976uz}. For the case of an instantaneous interaction, the light-front Bethe-Salpeter formalism 
automatically incorporates crossing. 

To summarize, the non-valence contribution to the instantaneous model's form factor is
\begin{equation} \label{NVI}
\delta F_I = (\delta J^+_1 + \delta J^+_2)/(1 - \zeta/2),
\end{equation}
and involves crossed interactions or equivalently non-wave function contributions.

\subsubsection{Comparison}
Now we compare the form factors for the cases of instantaneous and non-instantaneous interactions. 
%is obscured because the wave functions and their eigenvalues $M^2$ are different. 
In the instantaneous limit $\mu^2 \gg m^2, M^2$, however, we understand the behavior of the wave functions.  
The wave functions Eq.~\eqref{OBE} and Eq.~\eqref{nOBE} both become narrowly peaked about $x = 1/2$ in the large $\mu^2$ 
limit, \emph{cf} the behavior of $\dw(x,4m^2)$. 
%Already we have indicated one simplification for the comparison of form factors. 
Since we are investigating non-valence contributions to \emph{form factors}, we choose additionally 
to solve for the instantaneous wave function to first
order in $V - V_o$ (see Eq.~\ref{instW}) to put the wave functions on equal footing: i.e.~$\phi(x) \approx \psi(x)$, 
because the difference $V - V_o$ is presumed small. 
This has the efficacious consequence of producing identical leading-order terms \emph{cf} Eq.~\eqref{FFLO} and eliminates
the issue of normalization. 
%Indeed we will only be comparing non-valence contributions. 
The optimal choice of the instantaneous interaction is not under investigation here. So we shall simplify matters further
by choosing $E = 0$ in Eq.~\eqref{Vo}.
%, which results in a contact interaction\footnote{This can only be done with $\phi(x) = \psi(x)$ since otherwise
%the wave functions will not agree as $\mu^2 \to \infty$.}. 

\begin{table}
\begin{center}
	\begin{tabular}{|| l | r ||}
	\hline
	                $ \mu^2/m^2 $        &   $ 4 - M^2/m^2  $   \\ 
	\hline
			     $0.100$                  &   $6.57 \times 10^{-1}$     \\
	\hline
			     $0.316$                  &   $2.72 \times 10^{-1}$     \\	
	\hline
	                     $1.00$                   &   $6.34 \times 10^{-2}$     \\ 
        \hline
	                     $3.16$                   &   $8.79 \times 10^{-3}$     \\ 
	\hline
	                     $10.0$                   &   $9.57 \times 10^{-4}$     \\ 
	\hline
	                     $31.6$                   &   $4.81 \times 10^{-5}$     \\ 
	\hline
%	                     $78.3$                   &   $1\times10^{-7}$     \\ 
%	\hline	
	\end{tabular}
\end{center}
\caption{Numerical solution of the bound-state equation Eq.~\eqref{OBE} for various values of $\mu^2$.
The coupling constant $\alpha = 0.100$. }  
\label{bsvals}
\end{table}

\begin{figure}
\begin{center}
\epsfig{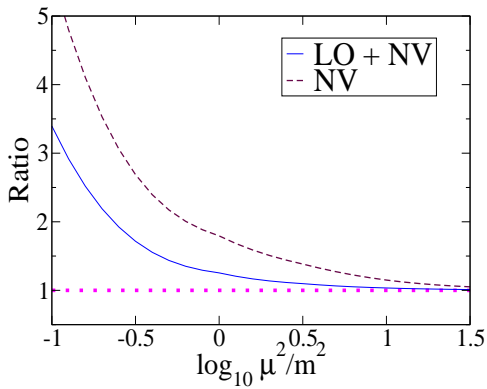}
\caption{Comparison of form factors in instantaneous and non-instantaneous cases. The ratio Eq.~\eqref{ratio} of form factors (LO + NV)
 is plotted versus $\log_{10}(\mu^2/m^2)$ at fixed $\zeta = 0.707$. 
Additionally the ratio Eq.~\eqref{ratioprime} of non-valence contributions to the form factor (NV) is plotted. 
For $\mu^2 \gg m^2, M^2$, the non-instantaneous contributions becomes approximately instantaneous.}
\label{it}
\end{center}
\end{figure}

For a few values of $\mu^2$, we solve for the wave function using Eq.~\eqref{OBE}. In Table \ref{bsvals}, we list the values of
$\mu^2$ used as well as the corresponding eigenvalue $M^2$ (all for the coupling constant $\alpha = 0.100$, where 
$\alpha = g^2 / 4 \pi m^4$). We then calculate the form
factors in the instantaneous and non-instantaneous cases. We arbitrarily choose $\zeta = 0.707$. 
In $(1+1)$ dimensions, this fixes $\Delta^2 = - \zeta ^2 M^2 / (1-\zeta)$. 
The ratio of these form factors---defined using Eqs.~(\ref{FFLO},\ref{NVNI},\ref{NVI}), namely
\begin{equation} \label{ratio}
(F_{LO} + \delta F_I)/(F_{LO} + \delta F_{NI} )
\end{equation} 
---is plotted versus $\log_{10}(\mu^2/m^2)$ 
in Figure \ref{it}.  The figure indicates the form factors are the same in the large $\mu^2$ limit. 
Since the leading-order contributions are identical, the Figure additionally shows 
the ratio of non-valence contributions to the form factor, namely
\begin{equation} \label{ratioprime}
\delta F_I / \delta F_{NI}.
\end{equation} 
This ratio too tends to one as $\mu^2$ becomes large, of course not as rapidly. 
Thus for finite $\mu^2$, we must add the correction $V - V_o$ in Eq.~\eqref{instW}
which results in higher Fock contributions present in the non-instantaneous case.

Lastly we remark that all of this is analytically clear: having eliminated different $\mu^2$ dependence hidden 
in $\psi$ and $\phi$, expressions for form factors in the instantaneous
and non-instantaneous cases are identical to leading order in $1/\mu^2$, i.e.~$\delta J^+_A, \delta J^+_B, \delta J^+_C
\sim 1/\mu^4$ while the remaining non-valence contributions $\delta J^+_D, \delta J^+_E$ match up 
with $\delta J^+_1$ and $\delta J^+_2$, respectively, to $1/\mu^2$. 

Of course we have demonstrated the replacement of non-wave function vertices only to first order in perturbation theory.  
Schematically, we have required
\begin{equation}
| V - V_o | \ll |V ( G - \Gt) V|.
\end{equation}
This is merely the condition that corrections to $V_o$ from finite $\mu^2$ are larger than second-order corrections
from the reduction scheme in Eq.~\eqref{instW}. 
%(which are second-order in $\alpha$ for the ladder model). 
Quantitatively this condition translates to $m/ 4 \pi \mu \ll \alpha$. 
If this condition is met, the leading corrections are all of the form $V - V_o$
and the instantaneous model becomes the ladder model (and non-wave function vertices disappear when calculating the form factor). 
When the condition is not met, but holds to $[V(G - \Gt)]^2 V$, say, we 
must work up to second order in the reduction scheme and use the second Born approximation. 
We do not pursue this lengthy endeavor here, since for the non-instantaneous case $\gamma(x|M^2) \propto \theta[x(1-x)]$ 
necessarily excludes non-wave function contributions in time-ordered perturbation theory.
%In fact, we expect to find a similar replacement of non-wave function vertices in favor of higher Fock 
%components. 
In fact at any order in time-ordered perturbation theory we expect to find a complete expression for the form 
factor in terms of Fock component overlaps devoid of non-wave function vertices, crossed interactions, \emph{etc}. 

\section{Summary} \label{sum}
Above we investigate current matrix elements in the light-front Bethe-Salpeter formalism.
First we present the issue of non-wave function vertices by taking up the common
assumptions of light-front constituent quark models. By calculating the form factor in frames where
$\zeta \neq 0$, Lorentz invariance mandates contribution from the vertex function's poles. 
Quark models which neglect these residues and postulate a form for the wave function are assuming not only a pole-free vertex,
but a vertex independent of light-front energy. These assumptions are very restrictive.

This leads us formally to investigate instantaneous and non-instantaneous contributions
to wave functions and form factors necessitating the reduction formalism Eq.~\eqref{324}.   
We provide an intuitive interpretation for the light-front reduction, namely it is a procedure 
for approximating the poles of the Bethe-Salpeter vertex function. This procedure consists of covariant 
iterations followed by an instantaneous approximation \emph{cf} Eq.~\eqref{interpretation}, where the
auxiliary Green's function $\Gt$ enables the instantaneous approximation, see Eq.~\eqref{key}.  
In order to calculate form factors in the reduction formalism, we construct the gauge invariant
current Eq.~\eqref{emvertex}.

Using the ladder model \eqref{ladder} and an instantaneous approximation \eqref{Vo} we compare 
the calculation of wave functions and form factors in the light-front reduction scheme. 
Calculation of form factors is dissimilar for the two cases. 
In the ladder model, which is non-instantaneous,
non-wave function vertices are excluded
%were reducible contributions and were hence removed by $\Gt$ 
in time-ordered perturbation theory. 
For the instantaneous model, however, contribution from crossed interactions
%non-wave function vertices 
is required. Moreover, 
these instantaneous contributions derived
are identical in form to non-wave function vertices used in constituent quark models. 
%where non-wave function vertices are used.  
As a crucial check on our results, we take the limit 
$\mu^2 \gg m^2, M^2$ for which 
the ladder model becomes approximately instantaneous. In this limit, calculation of the form factors for the two models is identical. 

The net result is an explicit proof, for $(1+1)$ dimensions, that non-wave function vertices are replaced by contributions form higher
Fock states if the interactions between particles are not instantaneous. 
We call this a replacement theorem. 
The analysis relies on general
features of the pole structure of the Bethe-Salpeter equation in light-front time-ordered perturbation theory. Therefore it is trivial to 
extend the theorem to $(3+1)$ dimensions. The real question which remains is whether or not the interaction between light quarks
can be approximated as instantaneous. We intend to learn how to use experimental data to answer this question.

\begin{acknowledgments}
We thank M.~Diehl for enthusiasm, questions and critical comments.
This work was funded by the U.~S.~Department of Energy, grant: 
DE-FG$03-97$ER$41014$.  
\end{acknowledgments}


\begin{thebibliography}{99}
%\cite{Dirac:1949cp}
\bibitem{Dirac:1949cp}
P.~A.~M.~Dirac,
%``Forms Of Relativistic Dynamics,''
Rev.\ Mod.\ Phys.\ {\bf 21}, 392 (1949).
%%CITATION = RMPHA,21,392;%%

%\cite{Leutwyler:1977vy}
\bibitem{Leutwyler:1977vy}
H.~Leutwyler and J.~Stern,
%``Relativistic Dynamics On A Null Plane,''
Annals Phys.\  {\bf 112}, 94 (1978).
%%CITATION = APNYA,112,94;%%

%\cite{Lepage:1980fj}
\bibitem{Lepage:1980fj}
G.~P.~Lepage and S.~J.~Brodsky,
%``Exclusive Processes In Perturbative Quantum Chromodynamics,''
Phys.\ Rev.\ D {\bf 22}, 2157 (1980).
%%CITATION = PHRVA,D22,2157;%%

%\cite{Brodsky:1997de}
\bibitem{Brodsky:1997de}
S.~J.~Brodsky, H.~C.~Pauli and S.~S.~Pinsky,
%``Quantum chromodynamics and other field theories on the light cone,''
Phys.\ Rept.\  {\bf 301}, 299 (1998).
%[arXiv:hep-ph/9705477].
%%CITATION = HEP-PH 9705477;%%

%\cite{Carbonell:1998rj}
\bibitem{Carbonell:1998rj}
J.~Carbonell, B.~Desplanques, V.~A.~Karmanov and J.~F.~Mathiot,
%``Explicitly covariant light-front dynamics and relativistic few-body  systems,''
Phys.\ Rept.\  {\bf 300}, 215 (1998).
%[arXiv:nucl-th/9804029].
%%CITATION = NUCL-TH 9804029;%%

%\cite{Miller:2000kv}
\bibitem{Miller:2000kv}
G.~A.~Miller,
%``Light front quantization: A technique for relativistic and realistic  nuclear physics,''
Prog.\ Part.\ Nucl.\ Phys.\  {\bf 45}, 83 (2000).
%[arXiv:nucl-th/0002059].
%%CITATION = NUCL-TH 0002059;%%

%\cite{Ligterink:1994tm}
\bibitem{Ligterink:1994tm}
N.~E.~Ligterink and B.~L.~Bakker,
%``Equivalence of light front and covariant field theory,''
Phys.\ Rev.\ D {\bf 52}, 5954 (1995);
%[arXiv:hep-ph/9412315].
%%CITATION = HEP-PH 9412315;%%
%\cite{Ligterink:wk}
%\bibitem{Ligterink:wk}
%N.~E.~Ligterink and B.~L.~Bakker,
%``Renormalization Of Light Front Hamiltonian Field Theory,''
%Phys.\ Rev.\ D {\bf 52}, 
5917 (1995).
%%CITATION = PHRVA,D52,5917;%%

%\cite{Sales:1999ec}
\bibitem{Sales:1999ec}
J.~H.~Sales, T.~Frederico, B.~V.~Carlson and P.~U.~Sauer,
%``Light-front Bethe-Salpeter equation,''
Phys.\ Rev.\ C {\bf 61}, 044003 (2000);
%[arXiv:nucl-th/9909029].
%%CITATION = NUCL-TH 9909029;%%
%\cite{Sales:2001gk}
%\bibitem{Sales:2001gk}
%J.~H.~Sales, T.~Frederico, B.~V.~Carlson and P.~U.~Sauer,
%``Renormalization of the ladder light-front Bethe-Salpeter equation in the Yukawa model,''
%Phys.\ Rev.\ C 
{\bf 63}, 064003 (2001).
%%CITATION = PHRVA,C63,064003;%%

%\cite{Tiburzi:2002mn}
\bibitem{Tiburzi:2002mn}
B.~C.~Tiburzi and G.~A.~Miller,
%``Light front Bethe-Salpeter equation applied to form factors,  generalized parton distributions and generalized distribution  amplitudes,''
hep-ph/0205109.
%%CITATION = HEP-PH 0205109;%%

%\cite{Ji:1998pc}
\bibitem{Ji:1998pc}
X.-D.~Ji,
%``Off-forward parton distributions,''
J.\ Phys.\ G {\bf 24}, 1181 (1998);
%[arXiv:hep-ph/9807358].
%%CITATION = HEP-PH 9807358;%%
%\cite{Radyushkin:2000uy}
%\bibitem{Radyushkin:2000uy}
A.~V.~Radyushkin,
%``Generalized parton distributions,''
hep-ph/0101225;
%%CITATION = HEP-PH 0101225;%%
%\cite{Goeke:2001tz}
%\bibitem{Goeke:2001tz}
K.~Goeke, M.~V.~Polyakov and M.~Vanderhaeghen,
%``Hard exclusive reactions and the structure of hadrons,''
Prog.\ Part.\ Nucl.\ Phys.\  {\bf 47}, 401 (2001).
%[arXiv:hep-ph/0106012].
%%CITATION = HEP-PH 0106012;%%

\bibitem{future}
B.~C.~Tiburzi and G.~A.~Miller, hep-ph/0210305.

%\cite{Brodsky:2001xy}
\bibitem{Brodsky:2001xy}
S.~J.~Brodsky, M.~Diehl and D.~S.~Hwang,
%``Light-cone wavefunction representation of deeply virtual Compton  scattering,''
Nucl.\ Phys.\ B {\bf 596}, 99 (2001);
%[hep-ph/0009254].
%%CITATION = HEP-PH 0009254;%%
%\cite{Diehl:2001xz}
%\bibitem{Diehl:2001xz}
M.~Diehl, T.~Feldmann, R.~Jakob and P.~Kroll,
%``The overlap representation of skewed quark and gluon distributions,''
Nucl.\ Phys.\ B {\bf 596}, 33 (2001).
%[hep-ph/0009255].
%%CITATION = HEP-PH 0009255;%%

%\cite{Bakker:2000rd}
\bibitem{Bakker:2000rd}
B.~L.~Bakker and C.-R.~Ji,
%``Disentangling intertwined embedded-states and spin effects in  light-front quantization,''
Phys.\ Rev.\ D {\bf 62}, 074014 (2000).
%[arXiv:hep-th/0003105].
%%CITATION = HEP-TH 0003105;%%

%\cite{Einhorn:1976uz}
\bibitem{Einhorn:1976uz}
M.~B.~Einhorn,
%``Form-Factors And Deep Inelastic Scattering In Two-Dimensional Quantum Chromodynamics,''
Phys.\ Rev.\ D {\bf 14}, 3451 (1976);
%%CITATION = PHRVA,D14,3451;%%
%\cite{Burkardt:2000uu}
%\bibitem{Burkardt:2000uu}
M.~Burkardt,
%``Off-forward parton distributions in 1+1 dimensional QCD,''
Phys.\ Rev.\ D {\bf 62}, 094003 (2000).
%[hep-ph/0005209].
%%CITATION = HEP-PH 0005209;%%

%\cite{Ji:2000fy}
\bibitem{Ji:2000fy}
C.-R.~Ji and H.-M.~Choi,
%``New effective treatment of the light-front nonvalence contribution in  timelike exclusive processes,''
Phys.\ Lett.\ B {\bf 513}, 330 (2001);
%[arXiv:hep-ph/0009281].
%%CITATION = HEP-PH 0009281;%%

%\cite{Tiburzi:2001je}
\bibitem{Tiburzi:2001je}
B.~C.~Tiburzi and G.~A.~Miller,
%``Exploring skewed parton distributions with two-body models on the light  front. II: Covariant Bethe-Salpeter approach,''
Phys.\ Rev.\ D {\bf 65}, 074009 (2002).
%arXiv:hep-ph/0109174.
%%CITATION = HEP-PH 0109174;%%

%\cite{Hwang:2001wu}
\bibitem{Hwang:2001wu}
C.~W.~Hwang,
%``Consistent treatment for valence and nonvalence configurations in  semileptonic weak decays,''
Phys.\ Lett.\ B {\bf 530}, 93 (2002).
%[arXiv:hep-ph/0108251].
%%CITATION = HEP-PH 0108251;%%

%\cite{deMelo:1997cb}
\bibitem{deMelo:1997cb}
J.~P.~B.~C.~de Melo, H.~W.~Naus and T.~Frederico,
%``Pion electromagnetic current in the light-cone formalism,''
Phys.\ Rev.\ C {\bf 59}, 2278 (1999);
%[arXiv:hep-ph/9710228].
%%CITATION = HEP-PH 9710228;%%
%\cite{deMelo:2002yq}
%\bibitem{deMelo:2002yq}
J.~P.~B.~C.~de Melo, T.~Frederico, E.~Pace and G.~Salm\`e,
%``Pair term in the electromagnetic current within the front-form  dynamics: Spin-0 case,''
Nucl.\ Phys.\ A {\bf 707}, 399 (2002).
%[arXiv:nucl-th/0205010].
%%CITATION = NUCL-TH 0205010;%%

%\cite{Jaus:zv}
\bibitem{Jaus:zv}
W.~Jaus,
%``Covariant Analysis Of The Light-Front Quark Model,''
Phys.\ Rev.\ D {\bf 60}, 054026 (1999).
%%CITATION = PHRVA,D60,054026;%%

%\cite{Demchuk:1995zx}
\bibitem{Demchuk:1995zx}
N.~B.~Demchuk, I.~L.~Grach, I.~M.~Narodetski and S.~Simula,
%``Heavy-to-heavy and heavy-to-light form factors for weak decays in the  light-front approach: Exclusive 0- to 0- case,''
Phys.\ Atom.\ Nucl.\  {\bf 59}, 2152 (1996);
%[Yad.\ Fiz.\  {\bf 59N12}, 2235 (1996)]
%[arXiv:hep-ph/9601369].
%%CITATION = HEP-PH 9601369;%%
%\cite{Grach:1996nz}
%\bibitem{Grach:1996nz}
I.~L.~Grach, I.~M.~Narodetsky and S.~Simula,
%``Weak decay form factors of heavy pseudoscalar mesons within a  light-front constituent quark model,''
Phys.\ Lett.\ B {\bf 385}, 317 (1996);
%[arXiv:hep-ph/9605349].
%%CITATION = HEP-PH 9605349;%%
%\cite{Cheung:1996qt}
%\bibitem{Cheung:1996qt}
C.~Y.~Cheung, C.~W.~Hwang and W.~M.~Zhang,
%``B $\to$ \pi l \nu Form Factors Calculated on the Light-Front,''
Z.\ Phys.\ C {\bf 75}, 657 (1997).
%[arXiv:hep-ph/9602309].
%%CITATION = HEP-PH 9602309;%%

%\cite{Schwinger:1951ex}
\bibitem{Schwinger:1951ex}
J.~S.~Schwinger,
%``On The Green's Functions Of Quantized Fields. 1,''
Proc.\ Nat.\ Acad.\ Sci.\  {\bf 37}, 452 (1951);
%%CITATION = PNASA,37,452;%%
%\cite{Gell-Mann:1951rw}
M.~Gell-Mann and F.~Low,
%``Bound States In Quantum Field Theory,''
Phys.\ Rev.\  {\bf 84}, 350 (1951);
%%CITATION = PHRVA,84,350;%%
%\cite{Salpeter:sz}
E.~E.~Salpeter and H.~A.~Bethe,
%``A Relativistic Equation For Bound State Problems,''
Phys.\ Rev.\  {\bf 84}, 1232 (1951).
%%CITATION = PHRVA,84,1232;%%

%\cite{Weinberg:1966jm}
\bibitem{Weinberg:1966jm}
S.~Weinberg,
%``Dynamics At Infinite Momentum,''
Phys.\ Rev.\  {\bf 150}, 1313 (1966).
%%CITATION = PHRVA,150,1313;%%

%\cite{Drell:1970km}
\bibitem{Drell:1970km}
S.~D.~Drell and T.~Yan,
%``Connection Of Elastic Electromagnetic Nucleon Form-Factors At Large Q**2 And Deep Inelastic Structure Functions Near Threshold,''
Phys.\ Rev.\ Lett.\ {\bf 24}, 181 (1970);
%%CITATION = PRLTA,24,181;%%
%\cite{West:1970av}
%\bibitem{West:1970av}
G.~B.~West,
%``Phenomenological Model For The Electromagnetic Structure Of The Proton,''
Phys.\ Rev.\ Lett.\ {\bf 24}, 1206 (1970).
%%CITATION = PRLTA,24,1206;%%

%\cite{Woloshyn:wm}
\bibitem{Woloshyn:wm}
R.~M.~Woloshyn and A.~D.~Jackson,
%``Comparison Of Three-Dimensional Relativistic Scattering Equations,''
Nucl.\ Phys.\ B {\bf 64}, 269 (1973).
%%CITATION = NUPHA,B64,269;%%
%\cite{Gross:bu}
\bibitem{Gross:bu}
F.~Gross and D.~O.~Riska,
%``Current Conservation And Interaction Currents In Relativistic Meson Theories,''
Phys.\ Rev.\ C {\bf 36}, 1928 (1987).
%%CITATION = PHRVA,C36,1928;%%

%\cite{Kvinikhidze:1998xn}
\bibitem{Kvinikhidze:1998xn}
A.~N.~Kvinikhidze and B.~Blankleider,
%``Gauging of equations method. I. Electromagnetic currents of three distinguishable particles,''
Phys.\ Rev.\ C {\bf 60}, 044003 (1999);
%[arXiv:nucl-th/9901001].
%%CITATION = NUCL-TH 9901001;%%
%\cite{Kvinikhidze:1999xp}
%\bibitem{Kvinikhidze:1999xp}
%A.~N.~Kvinikhidze and B.~Blankleider,
%``Gauging of equations method. II. Electromagnetic currents of three identical particles,''
%Phys.\ Rev.\ C {\bf 60}, 
044004 (1999).
%[arXiv:nucl-th/9901002].
%%CITATION = NUCL-TH 9901002;%%

%\cite{Sawicki:hs}
\bibitem{Sawicki:hs}
M.~Sawicki and L.~Mankiewicz,
%``Solvable Light Front Model Of A Relativistic Bound State In (1+1)-Dimensions,''
Phys.\ Rev.\ D {\bf 37} (1988) 421;
%%CITATION = PHRVA,D37,421;%%
%\cite{Mankiewicz:1989rb}
%\bibitem{Mankiewicz:1989rb}
L.~Mankiewicz and M.~Sawicki,
%``Solvable Light Front Model Of Electromagnetic Form-Factor Of The Relativistic Two Body Bound State In (1+1)-Dimension,''
Phys.\ Rev.\ D {\bf 40}, 3415 (1989).
%%CITATION = PHRVA,D40,3415;%%

\end{thebibliography}
\end{document}